\RequirePackage{fix-cm}
\documentclass[smallextended]{svjour3}       
\smartqed  
\usepackage{graphicx}
\begin{document}

\title{Extremal Limits and Kerr Spacetime 
}


\author{Partha Pratim Pradhan and Parthasarathi Majumdar}
\institute{\at Department of Physics\\
           Vivekananda Satabarshiki Mahavidyalaya \\
           Manikpara, Paschim Medinipur\\
            WestBengal-721513, India. \\
            \email{pppradhan77@gmail.com}
            \and 
            \at Department of Physics\\
           Ramakrishna Mission Vivekananda University \\
           Belur Math 711 202, India. \\
            \email{bhpartha@gmail.com}}
\date{Received: date / Revised version: date}

\maketitle

\begin{abstract}
The fact that one must evaluate the near-extremal and near-horizon limits of Kerr spacetime in a 
specific order, is shown to  lead to discontinuity in the extremal limit, such that this limiting 
spacetime differs nontrivially from the precisely extremal spacetime. This is established by first 
showing a discontinuity in
the extremal limit of the maximal analytic extension of the Kerr geometry, given by Carter. Next, we
examine the ISCO of the exactly extremal Kerr geometry and show that on the event horizon of the extremal
Kerr black hole, it coincides with the principal null geodesic generator of the horizon, having vanishing
energy and angular momentum. We find that there is no such ISCO in the near-extremal  geometry, thus 
garnering additional support for our primary contention. We relate this disparity between the two 
geometries to the lack of a trapping horizon in the extremal situation.

\end{abstract}

\section{Introduction}

The extremal limit for four dimensional Reissner Nordstr{\o}m (RN) black holes
has been extensively studied recently \cite{pppm,ddr,cjr} to probe whether the limit is continuous.
The definition of the Entropy Function \cite{wald} used to compare the `macroscopic' entropy of a class of
extremal black holes solutions of the supergravity limit of string theories,
to the `microscopic' entropy obtained from counting of string states
\cite{sen} requires a bifurcate horizon -an attribute that extremal
spacetimes do not possess. The trick of using a generic near-extremal
black hole spacetime to evaluate the entropy function, though widely
used in the literature, suffers from the pitfall that the extremal
limit may have subtle discontinuities \cite{ddr}, \cite{cjr}, and may
be different from the precisely extremal geometry. On our
part, it has been shown \cite{pppm} that there exists a class
of stable circular orbits on the event horizon of the extremal Reissner
Nordstr{\o}m spacetime which coincides with the principal null geodesic
generator of the horizon. The spacetime infinitesimally near the extremal
does not admit this class of geodesics.

In this paper we examine similar features of the {\it Kerr} spacetime :
whether in the extremal limit there exist geodesics on the horizon which
can be characterized as ISCOs, as compared with the precisely
extremal situation. Here, we employ Carter's \cite{c1,c2} maximal analytic extension
instead to investigate the continuity of the extremal limit vis-a-vis
the class of circular geodesics mentioned. While the use of conserved
scalars like `energy' and `angular momentum' of test particles, both massive
and massless, to study their geodesics, is usually deemed quite adequate in
any coordinate system, here, there is a word of caution: since the
geodesics in question lie {\it on} the horizon (and coincide with the null
geodesic generator \cite{pppm}, \cite{jacob}), the use of coordinate
charts smooth on the horizon is certainly preferable to the use of
charts which are not. Otherwise one may be led to conclusions which
may have additional subtle pitfalls. The main result that emerges from
our assay is what has been suspected earlier \cite{ddr} and pointed out recently
\cite{pppm} for the Reissner Nordstr{\o}m spacetime : {\it all aspects of the
extremal Kerr spacetime do not manifest themselves in the near-extremal limit}. In
particular, the class of geodesics close to the horizon in the extremal case
does not overlap with the class in the case that we are infinitesimally close to the
extremal.

The plan of the paper is as follows: in section \ref{mot}, we motivate our
paper by showing that the order in which the near-extremal and the near-horizon limits of the Kerr metric are taken, is
important, following a similar demonstration made in \cite{pppm} for the RN spacetime. The near horizon geometry of the
extremal black hole is {\it not} the same as the extremal limit of the near-horizon geometry of the generic non-extremal 
black hole. In section \ref{symm}, the Carter analytic extension of the Kerr spacetime is
discussed, on the axis of symmetry, and the difficulties of extracting the analytic extension of the extremal geometry 
from the extremal limit of the analytic extension of the
generic spacetime are elucidated. Equatorial circular
orbits are next considered in detail in section \ref{eco} with regard to their
stability, for timelike (outside the ergo-region) geodesics, and
compared for the extremal and the near-extremal geometries. Section \ref{trap} gives a discussion on the non-existence
of trapping surfaces in the
precisely extremal situation and contrasts this with the
near-extremal situation. We conclude our discussion in
section \ref{dis}. The Carter maximal extensions off the symmetry axis for extremal spacetimes are discussed in
an appendix\ref{app}.

\section{\label{mot}Motivation}

Consider the Kerr metric in Boyer-Lindquist coordinates,
\begin{eqnarray}
ds^2&=&-\frac{\Delta}{\rho^2} \, \left[dt-a\sin^2\theta d\phi \right]^2+\frac{\sin^2\theta}{\rho^2} \,
\left[(r^2+a^2) \,d\phi-a dt\right]^2
+\rho^2 \, \left[\frac{dr^2}{\Delta}+d\theta^2\right] ~.\label{nkm}
\end{eqnarray}
where
\begin{eqnarray}
a&\equiv&\frac{J}{M},\, \rho^2 \equiv r^2+a^2\cos^2\theta \nonumber\\
\Delta &\equiv& r^2-2Mr+a^2\equiv(r-r_{+})(r-r_{-})
\end{eqnarray}
For simplicity, taking the metric along the axis of symmetry
$(\theta=0)$, the Lorentzian 2-fold spanned by the coordinates $r,t$
has three possible geometries namely, near-horizon
geometry, near extremal geometry and the precisely extremal geometry. They
are defined in terms of two non-negative parameters  $\epsilon \equiv (r-r_1)/r_1 <<1$ and $\delta \equiv (r_+-r_-)/r_1 << 1$
where $r_1$ is the radius of the event horizon in the extremal case. In terms of these parameters,
\begin{eqnarray}
ds^2 = -{\epsilon (\epsilon + \delta) \over (1+\epsilon)} dt^2 + {(1+\epsilon)
\over \epsilon(\epsilon + \delta)} dr^2 ~.\label{near3}
\end{eqnarray}
which gives two different limiting geometries depending on the order in which the
limits $\delta \rightarrow 0$ and $\epsilon \rightarrow 0$ are taken.
First taking the extremal limit $\delta \rightarrow 0$ and then the
near-horizon limit, the {\it local} geometry is that of an $AdS_2$.
\begin{eqnarray}
ds^2 \simeq -\epsilon^2 dt^2 + {r_1^2 \over \epsilon^2} d\epsilon^2 ~.\label{ads2}
\end{eqnarray}

If, in contrast, the near horizon limit is taken before the extremal limit, one
obtains
\begin{eqnarray}
ds^2 \simeq - \epsilon \delta dt^2 + {r_1^2 \over \epsilon \delta } d\epsilon^2 ~.\label{nrh1}
\end{eqnarray}
which indicates that the local geometry is not an $AdS_2$ and the extremal limit is indeed now singular. What this establishes
is the subtlety that the near-horizon geometry of the extremal spacetime is not the same as the extremal limit of the near-horizon
geometry of the generic non-extremal spacetime, as advertised in the Introduction. Of course, for the latter case, the behavior
of the spacetime away from a bifurcation surface is what is being considered.

\section{\label{symm}Carter's Maximal Analytical Extension of Kerr Spacetime Along the Symmetric Axis}

The maximal analytic extension of the extremal Kerr spacetime along
the axis of symmetry was first reported by Carter \cite{c1}. Since, there is no discussion in the literature
about the  discontinuity of this extension at the extremal
limit $r_+ \rightarrow r_-$, we present here the complete maximal analytic extension following
Carter, showing that in the extremal limit $r_+ \rightarrow r_-$ it is
indeed discontinuous, necessitating a separate treatment.

\subsection{Non-Extremal Case:}

The  tortoise coordinate $r^\ast$ is given by
\begin{equation}
dr^\ast=\frac{(r^2+a^2)dr}{\Delta}=\frac{(r^2+a^2)dr}{(r-r_{-})(r-r_{+})}~.  \label{4.10}
\end{equation}
Integrating this equation, we obtain
\begin{equation}
F(r)=2r^\ast=2r+\kappa_{+}^{-1}\ln|r-r_{+}|
+\kappa_{-}^{-1}\ln|r-r_{-}| \label{4.30}
\end{equation}
with $\kappa_{\pm}^{-1} \equiv
\frac{2(r_{\pm}^2+a^2)}{(r_{\pm}-r_{\mp})}$ and where as usual
$r_{\pm}=M\pm\sqrt{M^2-a^2}$. $\kappa_{\pm}=$ is the surface gravity of the respective horizons. The outer horizon $r_{+}$ is
an event horizon and the inner horizon $r_{-}$ is a Cauchy horizon. $\kappa_{\pm}$ are both positive. $F(r)$ is monotonic in
each of the regions
\begin{eqnarray}
Region~ I: ~~~&&(r_{+}< r <\infty)\,\, \nonumber     \\
Region~ II:~~ &&(r_{-}< r< r_{+})\,\,  \nonumber    \\
Region~ III:~ &&(0< r <r_{-})\,\,     \label{effe1}
\end{eqnarray}
and blows up at the boundaries of the regions. Clearly, it is impossible to define a single coordinate patch which is
regular (in terms of geodetic completeness) over the entire spacetime.

Near the event horizon $r=r_{+}$, the tortoise coordinates is given by
\begin{equation}
r^\ast\approx\frac{1}{2\kappa_{+}}\ln|r-r_{+}| ~. \label{4.4}
\end{equation}
Here $r^\ast$  has logarithm dependence and is singular at
$r=r_{+}$.
Therefore, introducing double null coordinates $u \equiv r^{\ast}+t, v \equiv
r^{\ast}-t$, it is obvious that the event horizon $r=r_+$ occurs at $v+u=
-\infty$. The analytic extension of the Kerr spacetime is obtained by choosing in
the neighborhood of the event horizon the coordinate system
\begin{equation}
U^{+}=-\exp{(-\kappa_{+} u)},~~ V^{+}=\exp{(\kappa_{+} v)}\label{4.60}
\end{equation}

Therefore the metric becomes near $r=r_{+}$ along the axis of symmetry
\begin{eqnarray}
 ds^2 = -\frac{r_{+}r_{-}}{\kappa_{+}^2} \, \frac{\exp{(-2\kappa_{+}r})}{(r^2+a^2)}
      \, \left(\frac{r_{-}}{r-r_{-}}\right)^{\kappa_{+}/\kappa_{-}-1} dU^{+} dV^{+} \label{gmnr+}\nonumber
\end{eqnarray}

where

\begin{eqnarray}
 U^+ \, V^+=-\exp{(2\kappa_{+} r)} \, \left(\frac{r-r_{+}}{r_{+}}\right) \, \left(\frac{r-r_{-}}{r_{-}}\right)^{\frac{\kappa_{+}}{\kappa_{-}}} \label{uv1}
\end{eqnarray}

It is clear that $U^{+}=V^{+}=0$ corresponds to the bifurcation 2-sphere for the
generic Kerr spacetime. The coordinate patch used here is smooth in the neighbourhood
of the event horizon but {\it not} near the Cauchy horizon at $r=r_-$. Therefore, when
the two  horizons merge in the extremal limit, this coordinate patch is invalid. This conundrum shows up
in some metric coefficients in (\ref{gmnr+}) blowing up in the extremal limit $r_+ \rightarrow r_-$.

Similarly, a coordinate chart that is smooth across the Cauchy horizon can be constructed. The metric near $r=r_{-}$ along
the symmetry axis is
\begin{eqnarray}
 ds^2=-\frac{r_{+}r_{-}}{\kappa_{-}^2} \, \frac{\exp{(-2\kappa_{-}r})}{(r^2+a^2)} \,
      \left(\frac{r_{+}}{r-r_{+}}\right)^{\kappa_{-}/\kappa_{+}-1} dU^{-} dV^{-} \label{gmnr-}\nonumber
\end{eqnarray}
where
\begin{eqnarray}
 U^-  \, V^- &=& -\exp{(2\kappa_{-} r)} \, \left(\frac{r-r_{-}}{r_{-}}\right) \, \left(\frac{r-r_{+}}{r_{+}}\right)^{\frac{\kappa_{-}}{\kappa_{+}}} \label{uv--}
\end{eqnarray}

Once again, this coordinate chart is {\it not} a valid one near the event horizon,
and expectedly, the extended metric (\ref{gmnr-}) in the neighborhood of the
$r_-$ blows up in the extremal limit $r_{+} \rightarrow r_{-}$.

\subsection{Extremal Case:}

Thus we cannot  obtain the complete maximal analytic
extension along the axis of symmetry of the {\it extremal} Kerr
spacetime as a limiting case of the non-extremal Kerr spacetime;
the extremal case needs to be treated separately \cite{c1}.

The tortoise coordinate in this case is given by
\begin{equation}
r^\ast=\int\frac{(r^2+M^2)dr}{(r-M)^2}=r+ 2M \left[ \ln \left|r-M \right| -{M \over (r-M)} \right]  ~. \label{4.13}
\end{equation}
Near the horizon $r=M$ this has a leading pole-type singularity
\begin{equation}
r^\ast\approx\frac{2M^2}{(r-M)} \label{4.140}
\end{equation}
instead of a logarithmic one. Defining the double null coordinates $u$ and $v$
as $u \equiv r^\ast + t $, $v \equiv r^\ast - t $, the metric is given by
\begin{equation}
ds^2=\frac{(r-M)^2}{r^2+M^2}dudv  ~. \label{4.150}
\end{equation}
To determine the position of event horizon at a finite region in the coordinate chart, we
follow ref. \cite{c1} and introduce universal coordinates $U~,~V$ such that
\begin{eqnarray}
u = \tan U~, ~ v = \cot V~. \label{4.16}
\end{eqnarray}
This implies that
\begin{eqnarray}
\tan U + \cot V = 2 r^{\ast}~(U,V) . \label{4.17}
\end{eqnarray}
Therefore the extremal Kerr metric along the axis of symmetry in the Carter
coordinate system is given by
\begin{eqnarray}
ds^2 &=& - \frac{(r-M)^2}{r^2+M^2} \sec^2U \csc^2VdUdV ~.\label{4.18}
\end{eqnarray}

The geodesic orbits of the test particle on the 2-fold can be obtained by considering the spacetime on the symmetry axis.
Conserved quantities like test particle energy and angular momentum (both per unit mass) are defined in terms of Killing
vector fields corresponding to stationarity and axisymmetry of the Kerr spacetime. E.g., the energy is defined as
$\xi \cdot {\bf u} \equiv -{\cal E}$, where
$\bf u$ is the four velocity  of the test particle and ${\cal E}$ is the energy per unit mass of the test particle. The timelike
Killing vector denoted as $\xi\equiv \partial_t$ and its components are
$\xi^U=\cos^2 U~,~\xi^{V} = \sin^2 V$ . In the
next section we will be describe it in more detail. Now using the normalization
condition of four velocity and for circular orbit we get the part of the squared potential along the symmetry axis
\begin{eqnarray}
 {\cal E}^2={\cal V}_{squar}=-\frac{(r-M)^2}{r^2+M^2}\kappa
\end{eqnarray}
where $\kappa=-1$ for timelike geodesics and $\kappa=0$ for null geodesics. In other words, the ISCO turns into a null
geodesic on the horizon $r(U,V) = M$ with the energy vanishing as well. Thus, on the horizon, the only class of geodesics
that survives must be null with vanishing energy. Hence as $r(U,V)\rightarrow M$
\begin{eqnarray}
 {\cal E}^2={\cal V}_{squar}={\cal U}_{squar}=0
\end{eqnarray}
where ${\cal V}_{squar}$ and ${\cal U}_{squar}$ are the squared potentials corresponding to the timelike and null cases
respectively. This implies that in the extremal geometry, the geodesic on the horizon corresponds to the null geodesic generator.

\section{\label{eco} Equatorial Circular Orbits in Kerr spacetime: Doran/Ingoing Kerr Coordinates}

To compute the effective potential for circular orbits we choose nonsingular
coordinates like ingoing Kerr coordinates rather than the Boyer Lindquist coordinates. In this coordinate system  the
metric\cite{c1} can be written as
\begin{eqnarray}
ds^2 &=&-(1-\frac{2Mr}{\rho^2})dv^2+2dvdr+\rho^2 \, d\theta^2-2a\sin^2\theta dr d\tilde{\phi} \nonumber \\[4mm] &&
+\frac{1}{\rho^2} \, \left[(r^2+a^2)^2-\Delta a^2\sin^2\theta\right] \, \sin^2\theta d\tilde{\phi}^2 
-\frac{4aMr}{\rho^2} \, \sin^2\theta d\tilde{\phi}dv~.\label{inke}
\end{eqnarray}
Taking the transformations
\begin{eqnarray}
dv &=& dt+\frac{dr}{1+\sqrt{2Mr(r^2+a^2)}} \\
d\tilde{\phi} &=& d\phi+\frac{a}{r^{2}+a^{2}+\sqrt{2Mr(r^2+a^2)}}dr ~.\label{tran}
\end{eqnarray}
we obtain the Doran form of Kerr metric\cite{doran} which is as follows:
\begin{eqnarray}
ds^2 =-dt^2+\left[\sqrt{\frac{2Mr}{\rho^2}}(dt-a\sin^2{\theta} d\phi)
+\sqrt{\frac{\rho^2}{r^2+a^2}}dr\right]^2+\rho^2{d\theta}^2 + (r^2+a^2)\sin^2{\theta}{d\phi}^2~.\nonumber\\
\label{pg}
\end{eqnarray}

\subsection{\label{tcgh}Circular Geodesics in the Precisely Extremal Case}

On the equatorial plane $\theta=\pi/2$, the extremal Kerr metric in Doran coordinates is given by
\begin{eqnarray}
ds^2 &=& -(1-\frac{2M}{r})dt^2-\frac{4M^2}{r}dtd\phi+
2\sqrt{\frac{2Mr}{r^2+M^2}}dtdr-2M\sqrt{\frac{2Mr}{r^2+M^2}}drd\phi
+\frac{r^2}{r^2+M^2}dr^2 \nonumber \\[4mm] && +\left[r^2+M^2+\frac{2M^3}{r}\right]d\phi^2~.\label{epg1}
\end{eqnarray}
In Doran coordinates,  the spacetime allows the timelike Killing vector $\xi
\equiv \partial_t$, whose projection along the four velocity ${\bf u}$ (${\bf
u}^2=-1$ for timelike and ${\bf u}^2=0$ for null) of geodesics: $\xi \cdot{\bf
u} = -{\cal E}$, is conserved along such geodesics. There is also the `angular
momentum ' $L \equiv \zeta \cdot {\bf u}$ (where $\zeta
\equiv \partial_{\phi}$) which is similarly conserved.  Thus, in this
coordinate chart, ${\cal E}$  and $L$ can be expressed as
\begin{eqnarray}
{\cal E} &=& (1-\frac{2M}{r})u^t-\sqrt{\frac{2Mr}{r^2+M^2}}u^r+\frac{2M^2}{r}u^{\phi}\\
L &=& (r^2+M^2+\frac{2M^3}{r})u^{\phi}-M\sqrt{\frac{2Mr}{r^2+M^2}}u^r-\frac{2M^2}{r}u^t.
\end{eqnarray}

We can not solve $u^{t}$ and $u^{\phi}$ in terms of ${\cal E}$ and $L$ on the horizon because in that situation $L=2M{\cal E}$ with
\begin{eqnarray}
{\cal E}&=& -u^t+2Mu^{\phi} ~\label{engt}
\end{eqnarray}
where, on the null surface $r=M~,~u^r=0$. Because of this degeneracy, a nontrivial solution for $u^t~,~u^{\phi}$ ensues if
and only if ${\cal E}= 0 = L$. The norm of the four velocity at $r=M$ is thus
\begin{equation}
{\bf u.u}=(u^{t})^2+4M^2(u^{\phi})^2-4Mu^{\phi}u^{t}={\cal E}^2
\end{equation}
implying that this is a null geodesic generator of the horizon. This agrees with the foregoing analysis in the
Carter frame in the previous section. Precisely on the event horizon in the exactly extremal geometry, there is no geodesic 
with non-vanishing energy or angular momentum (per unit mass).

\subsection{Near Extremal Kerr Spacetime:}

Proceeding similarly for non-extremal Kerr spacetime in Doran coordinates, the energy and angular momentum per unit mass of 
the test particle are
\begin{eqnarray}
{\cal E} &=& (1-\frac{2M}{r})u^t-\sqrt{\frac{2Mr}{r^2+a^2}}u^r+\frac{2aM}{r}u^{\phi}~.\\
L &=& (r^2+a^2+\frac{2Ma^2}{r})u^{\phi}-a \sqrt{\frac{2Mr}{r^2+a^2}}u^r-\frac{2aM}{r}u^t
\end{eqnarray}
For circular orbits at $r=r_{0}$, $u^r=0$, so that
\begin{eqnarray}
u^{t} &=& \frac{1}{\Delta}[(r_{0}^2+a^2+\frac{2Ma^2}{r_{0}}){\cal E} -\frac{2aM}{r_{0}}L] ~.\label{utt}\\
u^{\phi} &=& \frac{1}{\Delta}[ (1-\frac{2M}{r_{0}})L+\frac{2aM}{r_{0}}{\cal E}] ~.\label{uf}
\end{eqnarray}
The squared norm of the four velocity
\begin{eqnarray}
{\bf u.u}
=-(1-\frac{2M}{r_{0}})(u^{t})^2-\frac{4aM}{r_{0}}u^{\phi}u^{t}+(r_{0}^2+a^2+\frac{2Ma^2}{r_{0}})
(u^{\phi})^2 ~\label{uum}
\end{eqnarray}
which reduces to
\begin{eqnarray}
{\bf u.u} &=& -{\cal E}u^{t}+Lu^{\phi}
\end{eqnarray}
 More explicitly
\begin{eqnarray}
{\bf u.u} &=& \frac{1}{\Delta}[(L^2-a^2{\cal E}^2)-r_{0}^2{\cal E}^2-\frac{2M}{r_{0}}(L-a{\cal E})^2] ~.\label{uuf}
\end{eqnarray}
Now if we take the extremal limit $a \rightarrow M$ and the near-horizon limit $\Delta \equiv (r_{0}-M)^2 \rightarrow 0$, one obtains,
\begin{eqnarray}
{\bf u.u} &=&  -\frac{(L-2M{\cal E})^2}{(r_{0}-M)^2} ~.\label{uufx1}
\end{eqnarray}
The proof that in the extremal limit the circular ISCO turns into a null geodesic generator must first demonstrate that
the four velocity of the geodesic has vanishing norm on the horizon. However, it is not clear that eq.(\ref{uufx1}) does 
lead to that, except for a specific order of limits mentioned above. Thus, in the extremal limit, we do have
$L=2M{\cal E}$, but to get ${\bf u.u} =0$, we must move slightly away from the horizon $r_0=M$; but this new 
hypersurface is not necessarily null, so the geodesic on it need not be null. On the other hand, if we move
away from extremality $L=2M{\cal E}$, then the near-horizon value of ${\bf u.u}$ diverges, and one can scarcely 
claim the existence of a null geodesic generator. There may be a way to evaluate the limits simultaneously, but
it is not clear how that establishes that the geodesic on the horizon at extremality is null. This conundrum is 
related to the issues mentioned in the Introduction, which motivate our primary contention that the extremal limit and the 
exactly extremal spacetimes have subtle disparities. It may be noted that our computational results in this subsection 
are consistent with those in ref. \cite{jacob}, as they are with those of the earlier paper \cite{hk}. However, rather 
than focus on the proper distance between the bifurcation sphere (nonexistent in the precisely extremal geometry) and 
the geodesic in question, we have chosen here to focus on the squared norm of the geodesic on the horizon in the 
extremal limit, and compare it with the behaviour in the exactly extremal situation. It is not enough to merely 
argue that the extremal limit exists, it is important to establish that an independent treatment of the exactly 
extremal case yields the same results as in the limiting situation. Since the precisely extremal situation has 
not been analyzed in \cite{jacob}, the comparison of the results there with those in the exactly extremal 
geometry cannot be made. In contrast, in this paper, we have already discussed the exactly extremal Kerr geometry 
in the last subsection, where ambiguities in evaluating limits seen in this subsection are not present.
This is yet another instance of the subtle disparity between the exactly extremal and the extremal limit
of a non-extremal Kerr spacetime, as we have contended, now observed within the Doran frame. In the next 
section, we reanalyze the extremal situation using Carter's maximal analytic extension of the extremal Kerr geometry.

\section{\label{cart} Carter's Maximal Analytic Extension of Extremal Kerr Spacetime:}

In section \ref{tcgh} we observed that both Doran and ingoing Kerr coordinates are better behaved at $r=M$ than the Boyer-Lindquist
coordinates; but they are not fully well behaved because \emph{``the outgoing coordinates  describe in a non-pathological
manner the ejection of particles outward from $r = 0$ through $r = 2M$; but their descriptions of in fall through $r=2M$  has
the same pathology as the description given by Schwarzschild coordinates. Similarly, the ingoing
coordinates  describe well the in fall of a particle through $r =2M$, but they
give a pathological description of outgoing trajectories..."}\cite{MTW}[Page-831]. This is exactly the same pathology of 
Boyer-Lindquist coordinates for the descriptions of Kerr spacetimes also. So here we will use Carter's coordinates/Universal 
like coordinates for the particular $r=M$ geodesics on the horizon. Carter's coordinates are explicitly derived in 
appendix \ref{app}. Without loss of generality we set $\dot{\theta}=0$ and $\theta=constant=\frac{\pi}{2}$ for the 
equatorial plane. Therefore from (\ref{mUVx}) the extremal Kerr metric on the equatorial plane can be written as
\begin{eqnarray}
ds^2 &=& {\cal A}\left(\sec^4U\, dU^2+\csc^4V\,dV^2 \right)+{\cal B}\sec^2U \, \csc^2VdUdV +{\cal C}\,(d\phi^{\star})^2 \nonumber \\[4mm]&&
+{\cal D}\left(\sec^2U\, dU+\csc^2V\, dV\right)d\phi^{\star} ~.\label{metex}
\end{eqnarray}
where
\begin{eqnarray}
{\cal A}&=&\frac{1}{8}(1-\frac{M}{r})^2 \left(\frac{r^2}{r^2+M^2}+\frac{1}{2}\right) \left(\frac{r^2-M^2}{r^2+M^2}\right)+\frac{(M^2-r^2)^2}{16M^2r^2} ~.\label{a}\\
{\cal B}&=&\frac{(M^2-r^2)^2}{8M^2r^2}-\frac{1}{2}(1-\frac{M}{r})^2 \, \left[\frac{r^4}
{(r^2+M^2)^2}+\frac{1}{4}\right] ~.\label{b}\\
{\cal C}&=&\frac{(r^2+M^2)^2}{r^2}-M^2 \, (1-\frac{M}{r})^2,\,
{\cal D}=\frac{1}{2}M \, (1-\frac{M}{r})^2-\frac{(M^4-r^4)}{2Mr^2} ~.\label{d}
\end{eqnarray}
For our simplicity we take $\phi$ instead of $\phi^{\star}$ in the subsequent analysis.
The spacetime metric (\ref{metex}) has a timelike isometry. The generator of this isometry is the Killing vector field ${\bf \xi}$ whose
projection along the 4-velocity ${\bf
u}$ of timelike geodesics:  $\xi \cdot {\bf u} = -{\cal E}$, is conserved along such geodesics. Now, ${\bf \xi}$ has non-vanishing
components $\xi^U,\xi^V$ which can be easily derived
from the fact that in the Schwarzschild coordinate basis ${\bf \xi} = {\bf \partial}_t$.
One obtains $\xi^U=\cos^2 U~,~\xi^{V} = \sin^2 V$. Thus, in this coordinate chart, ${\cal E}$ can be expressed as
\begin{eqnarray}
{\cal E}&=&-({\cal A}+{\cal B}/2) \, \left[\sec^2U \, u^U+\csc^2V \, u^V \right]+2{\cal D} \, u^{\phi}
\label{engm}
\end{eqnarray}
There is also  other isometry for rotational symmetry i.e. the `angular momentum'
$L \equiv \zeta \cdot {\bf u}$ (where $\zeta
\equiv \partial_{\phi}$) which is similarly conserved. It can be also expressed as
\begin{eqnarray}
L &=& {\cal D} \, \left[\sec^2U \, u^U+\csc^2V \, u^V\right]+{\cal C} \, u^{\phi}~.\label{mtld}
\end{eqnarray}
From the norm of four velocity
\begin{eqnarray}
u^2 &=& {\cal A} \, \left[\sec^4U \, (u^U)^2+\csc^4V \, (u^V)^2\right]+{\cal B} \, \sec^2U \, \csc^2V \, u^U \, u^V \nonumber \\[4mm] &&
+{\cal D} \, \left[\sec^2U \, u^U +\csc^2V \, u^V \right] \, u^{\phi}+{\cal C} \, (u^{\phi})^2~. \label{u2u2}
\end{eqnarray}
Now, for {\it circular} geodesics; the radial component of the 4-velocity vanishes:
$u^r=0$. Therefore, this translates into $r_U u^U + r_V u^V =0$ where $r_U \equiv \partial r / \partial U$ etc.
These derivatives of $r(U,V)$ can be calculated from equation (\ref{xtan}), so that one obtain for circular geodesics
\begin{eqnarray}
{u^U \over u^V} &=& \cos^2 U  \,  \csc^2 V ~. \label{uux}\\
{\cal E} &=& -\left [2({\cal A}+{\cal B}/2) \, \sec^2U \, u^U+2{\cal D}\, u^{\phi}\right]~.\label{enm}\\
L &=& 2{\cal D} \, \sec^2U \, u^U+{\cal C} \, u^{\phi} ~.\label{metld}
\end{eqnarray}
Solving equations (\ref{enm}), (\ref{metld}) one obtains
\begin{eqnarray}
u^U \, \sec^2U &=& \frac{2{\cal D} \, L+{\cal C} \, {\cal E}}{4{\cal D}^2-2{\cal C}({\cal A}+{\cal B}/2)} ~. \label{dotu}\\
u^{\phi} &=& \frac{L({\cal A}+{\cal B}/2)+{\cal D}{\cal E}}{{\cal C}({\cal A}+{\cal B}/2)-4{\cal D}^2} ~. \label{ph}
\end{eqnarray}
From (\ref{u2u2})
\begin{eqnarray}
{\bf u}^2 &=& -(u^U \, \sec^2U) \,  {\cal E} +{\cal C} \, (u^{\phi})^2 .~ \label{frvt}
\end{eqnarray}
Now we would like to see what happens for the peculiar geodesics $r=M$ in this fully well behaved coordinates. Does it coincide 
with the null generators of the horizon?  Note that in this coordinates the horizon is at  $r(U,V)=M$, $U=\pi/2$ and for circular
geodesics on the  future horizon
\begin{eqnarray}
\cal E &\rightarrow& 0 ~.\nonumber\\
{u^U \over u^V} &=& \cos^2 U \csc^2 V \rightarrow 0  ~.\label{limt1}
\end{eqnarray}
and the norm of the four velocity may be defined as
\begin{equation}
 {\bf u.u}=\frac{L^2}{4M^2}
\end{equation}
Once again, as in the Doran frame for the exactly extremal Kerr geometry, the geodesic on the horizon will be a null geodesic 
generator only if $L=0$. This also implies that the energy ${\cal E}$ must also vanish for the geodesic on the horizon.

Alternatively for timelike circular geodesics ${\bf u}^2=-1$, Using (\ref{dotu},~\ref{ph}) one  obtain
the energy equation for timelike circular geodesics as
\begin{eqnarray}
\alpha \, {\cal E}^2+\beta \, \cal{E}+\gamma &=& 0  ~.\label{e2}
\end{eqnarray}
where
\begin{eqnarray}
\alpha &=& G\,{\cal C}\,{\cal D}^2-{\cal C}\,{\cal H}^2 \nonumber\\
\beta &=& 2{\cal C}\,{\cal D}GL\,({\cal A}+{\cal B}/2)-2{\cal D}L{\cal H}^2 \nonumber\\
\gamma &=& {\cal C}\,GL^2 \, ({\cal A}+{\cal B}/2)^2-G \, {\cal H}^2 \nonumber\\
G &=& 4{\cal D}^2-2{\cal C} \, ({\cal A}+{\cal B}/2) \nonumber\\
{\cal H} &=& {\cal C} \, ({\cal A}+{\cal B}/2)-4{\cal D}^2  ~.\label{gh0}
\end{eqnarray}
Therefore the effective potential for timelike circular geodesics may be written as
\begin{eqnarray}
{\cal E} = ({\cal V}_{eff})_{Horizon} = \frac{-\beta+\sqrt{{{\beta}^2-4\alpha\gamma}}}{2\alpha}
~.\label{vef}
\end{eqnarray}
Similarly the effective potential for null circular geodesics can be written as
\begin{eqnarray}
{\cal E} = ({\cal U}_{eff})_{Horizon} = \frac{-\beta+\sqrt{{{\beta}^2-4\alpha\gamma_{0}}}}{2\alpha} ~.\label{uef}
\end{eqnarray}
where
\begin{eqnarray}
\gamma_{0} &=& {\cal C}\,GL^2\,({\cal A}+{\cal B}/2)^2~.\label{gama0}
\end{eqnarray}
It shows that the future horizon of the spacetime is given by $U=\pi/2$ with $V$ arbitrary: in other words $r(\pi/2, V)=M$. One
can compute the derivatives in the equations
(\ref{xtan}); it turns out that $r_U (\pi/2,V)= 2M^2$, while the other derivative of
$r$ is regular on the horizon. Now on the future horizon
\begin{eqnarray}
{\cal A}\rightarrow0,~~{\cal B}\rightarrow 0,~~{\cal C}\rightarrow 4M^2,~~ {\cal D}\rightarrow 0 \\
{\alpha}\rightarrow 0,~~{\beta} \rightarrow 0,~~{\gamma} \rightarrow 0,~~G \rightarrow 0,~~ {\cal H} \rightarrow 0 \label{abcd}
\end{eqnarray}
\begin{eqnarray}
\cal E &\rightarrow& 0 ,\,
{u^U \over u^V} = \cos^2 U \csc^2 V \rightarrow 0  ~.\label{limt}
\end{eqnarray}
which implies  $u^U\rightarrow 0$ and $u^V\rightarrow \infty$ on the
horizon. It follows that $L$ is a finite quantity which is
vanishing on the horizon. It  is also further observed that timelike circular geodesics and
null circular geodesics \emph{coalesce into a zero energy trajectory} (as in the RN case \cite{pppm})
\begin{eqnarray}
{\cal E} = ({\cal V}_{eff})_{Horizon}=({\cal U}_{eff})_{Horizon}\rightarrow 0  ~.\label{zeroE}
\end{eqnarray}
Thus, the geodesic on the horizon  must coincide with the principal null geodesic generator. The existence of a
timelike circular orbit turning into the null geodesic generator on the event horizon is a peculiar feature of exactly extremal 
Kerr spacetime.

Another view of this discontinuity is gleaned from the absence of outer trapped surfaces within the horizon in the extremal 
geometry in contrast to a more generic situation, as we now discuss.

\section{\label{trap}Absence of Trapped Surfaces in Extremal Kerr Spacetime:}
In a most general spacetime $(M,g_{\mu\nu})$  with the metric $g_{\mu \nu}$ having
signature $(-+++)$, one can define two future directed null vectors $l^{\mu}$ and
$n^{\mu}$ whose expansion scalars are given by
\begin{eqnarray}
\theta_{(l)}=q^{\mu\nu}\nabla_{\mu}l_{\nu}, \,\,\,
\theta_{(n)}= q^{\mu\nu}\nabla_{\mu}n_{\nu} ~. \label{expa}
\end{eqnarray}
where $q_{\mu\nu}=g_{\mu\nu}+l_{\mu}n_{\nu}+n_{\mu}l_{\nu}$ is the metric
induced by $g_{\mu \nu}$ on the two dimensional spacelike surface formed by
spatial foliation of the null hypersurface generated by $l^{\mu}$ and $n^{\mu}$.

Then (i) a two dimensional spacelike surface S is said to be a {\it trapped} surface if both $\theta_{(l)}<0$ and
$\theta_{(n)}<0$; (ii) S is to be {\it marginally trapped} surface if one of two null
expansions vanish i.e. $\theta_{(l)}=0$ or $\theta_{(n)}=0$. The null vectors
for non-extremal Kerr black hole are given by
\begin{eqnarray}
l^{\mu}&=& \frac{1}{\Delta}(r^2+a^2,-\Delta,0,a), \,\,\,n^{\mu}=\frac{1}{2r^2}(r^2+a^2,\Delta,0,a)\label{7.2}\\
l_{\mu}&=& \frac{1}{\Delta}(-\Delta,-(r^2+a^2),0,a\Delta), \,\,\, n_{\mu}=\frac{1}{2r^2}(-\Delta,r^2+a^2,0,a\Delta)\label{7.3}
\end{eqnarray}
where $\Delta=(r-r_{+})(r-r_{-})$ and $r_{\pm}=M\pm\sqrt{M^2-a^2}$.
The null vectors satisfy the following conditions :
\begin{equation}
l^{\mu}n_{\mu}=-1,\,\, l^{\mu}l_{\mu}=0, \,\, n^{\mu}n_{\mu}=0\label{7.4}
\end{equation}
Using (\ref{expa}), one obtains
\begin{equation}
\theta_{(l)}=-\frac{2}{r}, \,\, \theta_{(n)}=\frac{(r-r_{+})(r-r_{-})}{r^3}\label{7.5}
\end{equation}
In the region $(r_{-}< r< r_{+})$,  $\theta_{(l)}<0$ and
$\theta_{(n)}<0$. This implies that trapped surfaces exist for non extreme Kerr
black hole in this region. In contrast, for the extreme Kerr black hole
\begin{equation}
\theta_{(l)}=-\frac{2}{r}, \,\, \theta_{(n)}=\frac{(r-M)^2}{r^3}\label{7.6}
\end{equation}
Here inside or outside extremal horizon $r<M$ or $r>M$, $\theta_{(l)}<0$ and
$\theta_{(n)}>0$.
This implies that there are no trapped surfaces for extremal Kerr black hole
beyond the event horizon .

\section{\label{dis}Discussion:}

The study reveals the disparity between precisely extremely
and nearly extremely geometry manifested in the fact that the near-extremal and near-horizon {\it limits} do {\it not}
commute, and also the singular nature of the extremal limit of Carter's maximal analytic extension of a
generic Kerr geometry. We also showed that to study the geodesics close to the horizon, one must first go to the
{\it precisely} extremal geometry, before considering geodesics (near or on) the horizon by using Carter's frame which is 
well-behaved on the horizon. While Doran and ingoing-Kerr coordinates can also be used to reach the same conclusion, this 
is arrived at in the latter frames only through careful limiting procedures.

Another feature of our work is that using Carter's frame the direct ISCO in extremal Kerr spacetime, which lies {\it on} the 
event horizon, coincides with the principal null geodesic generator; such an ISCO is non-existent in the near-extremal geometry.

We have compared the results here with that in the work of Jacobson \cite{jacob} where it is also inferred that the ISCO on the 
horizon in the extremal case coincides with the null geodesic generator. For the near-extremal geometry, our results for the 
energy and angular momentum per unit mass agree with those in this work, and also with the earlier results of \cite{hk} 
(which is, surprisingly, not referred to in \cite{jacob}). However, the demonstration that the norm of the 4-velocity of 
the geodesic on the horizon must vanish in the extremal limit poses some challenging manipulations involving limits. In 
\cite{jacob}, a way around this problem has been sought by considering the proper spatial separation between the geodesic
in question, and the bifurcation sphere which does indeed exist in the near-extremal geometry. However, as emphasized in 
section IVB, it is not enough to argue that the extremal limit of a few quantities like the energy and momentum exist and 
are non-zero. One must analyze separately 
the exactly extremal geoemtry and compare the results with those in the extremal limit to exhibit the absence of subtleties
in that limit. This has {\it not} been done in \cite{jacob}, since the exactly extremal situation has not been considered
there. Indeed, for this geometry, there is no bifurcation sphere, and therefore arguments involving the proper distance 
between the geodesic and the bifurcation sphere cannot be made in this case. Rather, as we have unambiguously demonstrated 
in sections IVA and V, for the exactly extremal geometry, the geodesic on the horizon has vanishing norm provided the energy
and angular momentum per unit mass vanish. The demonstrations in this case are far more straightforward, requiring no subtle
manipulations of limits.

In sum, we hope to have persuaded the reader that the extremal limit of a generic non-extremal Kerr spacetime has subtle 
disparities from the precisely extremal situation. The evidence presented in favour of our contention is threefold : first 
of all, it is the absence of a unique maximal analytic extension of the non-extremal Kerr spacetime covering both the event
and Cauchy horizons, thus leading to a divergent metric in the extremal limit of such an extension. In other words, the 
maximal analytic extension of the extremal Kerr goemetry had to be worked out separately, rather than by a limiting 
procedure on the generic spacetime. Secondly, the existence of an ISCO which turns into the null geodesic generator 
of the horizon in the extremal case, with vanishing energy and angular momentum per unit mass. In the near-extremal 
situation, this demonstration is complicated because to show that the norm of the geodesic vanishes on the horizon 
requires careful handling of limits. Finally, our demonstration of the 
absence of trapped surfaces in the extremal spacetime, which tallies well with the other properties of the extremal 
spacetime noted here.

What we have apparently established has implications for extant approaches to computing the Wald entropy function for 
extremal black holes. It is not clear that current assays in this direction actually compute extremal black hole 
entropy; rather, the results are most likely for some sort of entanglement entropy of ambient matter in the field 
of such black holes.
\section{\label{app} Appendix : Maximal Analytical Extension of Kerr Spacetime (Off Axis of Symmetry)}
The maximal analytic extension of the non-extremal Kerr spacetime along
the off axis of symmetry was first reported by Carter \cite{c1}. As we showed for symmetry axis
the analytic extension is not continuous at the extremal limit $r_+ \rightarrow r_-$ in section \ref{symm},
here we implemented  for the off axis symmetry
and only derive the extremal case. For non-extremal case see Carter's \cite{c1} paper.

The complete analytic extension  of the {\it extremal} Kerr spacetime thus
cannot be obtained as a limiting case of the non-extremal geometry as previously, one needs to treat
the extremal case separately \cite{c1}. Defining the double null coordinates and angular coordinates are
\begin{eqnarray}
du+dv=2\frac{r^2+M^2}{\Delta}dr,\,
d\phi+d\tilde{\phi}=2\frac{M}{\Delta}dr ~.\label{dudvx}
\end{eqnarray}
Define ignorable angle coordinates $\phi^{\star}$ given by
\begin{equation}
2d\phi^{\star}=d\phi-d\tilde{\phi}-\frac{(du-dv)}{2M}~,  \label{kdfix}
\end{equation}
which are constant on the null generator of the horizon at $r=M$. The metric
is thus given by
\begin{eqnarray}
ds^2 &=& \frac{\Delta}{8\rho^2} \, \left[\frac{\rho^2}{r^2+M^2}+\frac{\rho_{\star}^2}{2M^2} \right]
\, \frac{(r^2-M^2)\sin^2\theta}{(r^2+M^2)}(du^2+dv^2)+\rho^2 \, d\theta^2 \nonumber \\
&+& \frac{\Delta}{2\rho^2} \, \left[\frac{\rho^4}{(r^2+M^2)^2}+\frac{\rho_{\star}^4}  {4M^4} \right] du dv
-\frac{\Delta M\sin^2\theta}{\rho^2} \,\left[M\sin^2\theta d\phi^{\star}-\frac{\rho_{\star}^2}{2M^2}(du-dv) \right] ~d\phi^{\star} \nonumber \\
&+& \frac{\sin^2\theta}{\rho^2} \, [ \frac{(M^2-r^2)}{4M}(du-dv)-(r^2+M^2)d\phi^{\star} ]^2 ~,\label{dnullx}
\end{eqnarray}
where $\rho_{\star}^2=M^2(1+\cos^2\theta)$, $\Delta=(r-M)^2$,
$\rho^2=(r^2+M^2\cos^2\theta)$. $r$ is now defined implicitly
as a function of $u$ and  $v$ by
\begin{equation}
F(r)=u+v=2r^\ast ~,\label{fuvx}
\end{equation}
where $r^\ast$ is called 'tortoise' coordinate, determined by
\begin{equation}
dr^\ast=\frac{(r^2+M^2)dr}{\Delta}=\frac{(r^2+M^2)dr}{(r-M)^2}~.  \label{rstarx}
\end{equation}
Integrating (\ref{rstarx}) yields
\begin{equation}
r^\ast=\int\frac{(r^2+M^2)dr}{(r-M)^2}=r+ 2M \left[ \ln \left|r-M \right| -{M \over 2(r-M)} \right]  ~. \label{torex}
\end{equation}
Near the horizon $r=M$ this has a leading pole-type singularity
\begin{equation}
r^\ast\approx\frac{M^2}{(r-M)} \label{rex}
\end{equation}
instead of a logarithmic one. To locate the event horizon at a finite region in
the coordinate chart, we follow ref. \cite{c1} and introduce null coordinates
$U~,~V$ such that
\begin{eqnarray}
u = \tan U~, ~ v = \cot V~. \label{uvx}
\end{eqnarray}
This implies that
\begin{eqnarray}
\tan U + \cot V = 2 r^{\ast}(U,V) ~. \label{xtan}
\end{eqnarray}
Therefore the complete extremal Kerr metric  in ($U~,V~,\theta~,\phi^\star$) is given by

\begin{eqnarray}
ds^2 &=& \frac{\Delta}{8\rho^2} \, \left[\frac{\rho^2}{r^2+M^2}+\frac{\rho_{\star}^2}{2M^2} \right] \,
\frac{(r^2-M^2)\sin^2\theta}{(r^2+M^2)} \, \left(\sec^4U dU^2+\csc^4VdV^2 \right )
+\rho^2 \, d\theta^2\nonumber \\[4mm] &&
-\frac{\Delta}{2\rho^2} \, \left[\frac{\rho^4}{(r^2+M^2)^2}+\frac{\rho_{\star}^4}
{4M^4}\right] \, \sec^2U \, \csc^2V \, dUdV\nonumber \\[4mm] &&
-\frac{\Delta M\sin^2\theta}{\rho^2} \left[M\sin^2\theta d\phi^{\star}
-\frac{\rho_{\star}^2}{2M^2}(\sec^2UdU+\csc^2VdV) \right] \, d\phi^{\star}
\nonumber \\[4mm] &&
+\frac{\sin^2\theta}{\rho^2}\left[\frac{(M^2-r^2)}{4M}\left(\sec^2U\,dU+\csc^2V \,dV \right)
-(r^2+M^2)d\phi^{\star}\right]^2 ~.\label{mUVx}
\end{eqnarray}
It can be easily checked  that in the limit $\theta=0$, we obtain the metric (\ref{4.18}) for symmetry axis.

\end{document}